# The quantization of energy in the harmonic-oscillator potential: Power series solution

Omer Sise

Afyon Kocatepe University, Science and Arts Faculty, Dept. of Physics, 03200, Afyonkarahisar, Turkey

**Abstract:**

We write a computer program that uses the recursion relation to calculate wave function in the harmonic-oscillator potential for specified values of $E/h\nu \pm 0.001$ containing only even numbers of $v$ (0,2,4,...). In this work, differential equations are solved by power series methods, i.e. the Schrodinger equation stationary state wavefunction of the harmonic-oscillator. This technique is applied to obtain the wavefunction of the quantum harmonic oscillator and is at a more sophisticated level than elsewhere in the course of quantum physics. For the able student this can be a worthwhile extension to the work on the harmonic-oscillator.

## 1. Introduction

The harmonic oscillator is an important system in the study of physical phenomena in both classical and quantum mechanics. It is well known that the harmonic-oscillator stationary-state energy levels are equally spaced, and are expressed by

$$E = (v + \frac{1}{2})h\nu \quad v = 0,1,2,\ldots$$

so that the energy is quantized in units of $h\nu$. For values of E that differ from this equation, $\psi$ is not quadratically integrable [1,2]. It is this requirement that the wavefunction not diverge so it can be normalized that yields energy quantization. There are many books that provides a uniquely accessible and thorough introduction to the quantum harmonic-oscillator for undergraduates [2-5].

In this paper, we study the quantization of the energy in the harmonic oscillator potential using power series solution. We write a computer program that uses the recursion relation to calculate wave function versus $x$, one-dimensional problem, for specified values of $E/h\nu \pm 0.001$ containing only even numbers of $v$ (0,2,4,...). It has been clearly shown that the cases of $E/h\nu \pm 0.001$ go to $\mp \infty$ at a particular value of $x$.

There are two procedures available for solving the differential equation of harmonic oscillator problems, ladder operator procedure and the Frobenius or series solution method. Both methods give exactly the same results. However, the solution of Schrödinger equation in power series gives a much more complete picture of the quantum-mechanical treatment of the harmonic oscillator which may be applied to the vibrations of molecular bonds and has many other applications in quantum physics and field theory.

## 2. Power series solution

The Schrödinger equation for the harmonic oscillator is given by

$$-\frac{\hbar^2}{2m}\frac{d^2\psi(x)}{dx^2} + \alpha x^2 \psi(x) = E\psi(x)$$

where $\alpha = m\omega^2/2$. We consider that the potential V(x) is approximated as a quadratic function of the bond displacement x expanded about the point at which V(x) is minimum. Then the resulting harmonic-oscillator equation can be solved exactly. Because the potential V(x) grows without bound as x approaches $\mp\infty$, only bound state solutions exist for this model problem; that is, the motion is confined by the nature of the potential, so no continuum states exist.

In solving the Schrödinger equation for this potential, the large-x behavior is first examined. The general solution to this equation is then taken to be of the form

$$\psi = e^{-\alpha x^2/2} \sum_{n=0,1,2,...}^{\infty} c_n x^n$$

where $c_n$ are coefficients to be determined. Substituting this expression into the full equation generates a set of recursion equations for the $c_n$ amplitudes. As in the solution of the hydrogen-like radial equation, the series described by these coefficients is divergent unless the energy E happens to equal specific values. The recursion equation is

$$\frac{c_{n+2}}{c_n} = \frac{\alpha[1+2n-2(E/hv)]}{(n+1)(n+2)}$$

where we made the substitution $mE\hbar^{-2} = E/hv$. Given $c_0$ it enables us (in principle) generate $c_2$, $c_4$, $c_6$, ... and given $c_1$ it generates $c_3$, $c_5$, $c_7$, .... The general solution of the Schrödinger equation is a linear combination of these two independent solutions. We now must see if the boundary conditions on the wave function lead to any restriction on the solution.

We have as a solution a power series containing only even powers of $x$ (n=0,2,4,...,2$\ell$, and $\ell$ =0,1,2,3,..., $\ell$ )

$$\psi = e^{-\alpha x^2/2} \sum_{n=0,2,4,...}^{\infty} c_n x^n = e^{-\alpha x^2/2} \sum_{\ell=0,1,2,...}^{\infty} c_n x^{2\ell}$$

If we rewrite this equation, we get

$$\psi = e^{-\left(\alpha^{1/2}x\right)^2/2}\left[c_0 x^0 + c_2 x^2 + c_4 x^4 + c_6 x^6 + ... + c_{2\ell} x^{2\ell}\right]$$

To compute $\psi/c_0$ we divided two sides of the equation by $c_0$

$$\psi/c_0 = e^{-\left(\alpha^{1/2}x\right)^2/2}\left[1 + \frac{c_2}{c_0}x^2 + \frac{c_4}{c_0}x^4 + \frac{c_6}{c_0}x^6 + ... + \frac{c_{2\ell}}{c_0}x^{2\ell}\right]$$

Also, we have

$$\frac{c_{2\ell+2}}{c_\ell} = \frac{\alpha[1+4\ell-2(E/hv)]}{(2\ell+1)(2\ell+2)} = \alpha A_\ell$$

$$A_\ell = \frac{[1+4\ell-2(E/hv)]}{(2\ell+1)(2\ell+2)}$$

We can change the variable $\ell$ in the recursion relation, and then calculate $c_\ell / c_0$.

$$\ell = 0 \qquad \frac{c_2}{c_0} = \alpha A_0$$

$$\ell = 1 \qquad \frac{c_4}{c_0} = \alpha^2 A_0 A_1$$

$$\ell = 2 \qquad \frac{c_6}{c_0} = \alpha^3 A_0 A_1 A_2$$

$$\cdots \qquad \cdots$$

$$\ell = \ell \qquad \frac{c_{2\ell+2}}{c_0} = \alpha^{\ell+1} A_0 A_1 A_2 \cdots A_\ell$$

Thus, we get

$$\psi / c_0 = e^{-(\alpha^{1/2} x)^2 / 2} \left[ 1 + \alpha A_0 x^2 + \alpha^2 A_0 A_1 x^4 + \alpha^3 A_0 A_1 A_2 x^6 + \cdots + \alpha^{\ell-1} A_0 A_1 A_2 \cdots A_\ell x^{2\ell} \right]$$

Using $\alpha^{1/2} x$ term (to eliminate $\alpha$), we have the final equation

$$\psi / c_0 = e^{-(\alpha^{1/2} x)^2 / 2} \left[ 1 + A_0 (\alpha^{1/2} x)^2 + A_0 A_1 (\alpha^{1/2} x)^4 + A_0 A_1 A_2 (\alpha^{1/2} x)^6 + \cdots + A_0 A_1 A_2 \cdots A_\ell (\alpha^{1/2} x)^{2\ell} \right]$$

### 3. Computer Programs

We write a computer program that uses the recursion relation to calculate $\psi / c_0$ versus $\alpha^{1/2} x$ values from -6 to 6 in increments of 0.1 for specified values of $E/h\nu \pm 0.001$. For example, the lowest value of the energy or zero-point energy is $h\nu/2$ (classically, the lowest energy for an oscillator is zero), and then we calculate $\psi / c_0$ for $E/h\nu$ =0.499, 0.500, and 0.501. Our program written in C-language can easily switch between the various $E/h\nu$ choices. This program includes a test to stop adding terms in the infinite series when the last calculated term is sufficiently small.

To see how the infinite series behave for large $x$, we examine the ratio of successive coefficients ($A_\ell$) in each series. In Fig. 1, we show the products of coefficients as a function of $\ell$ for $E/h\nu$ =0.499. It should be noted that $E/h\nu$ =0.501 gives the same results in negative direction, but $E/h\nu$ =0.500 gives always zero due to the recursion relation for $\ell$ =0 (in this case A$_0$=0, and all products of the coefficients are zero).

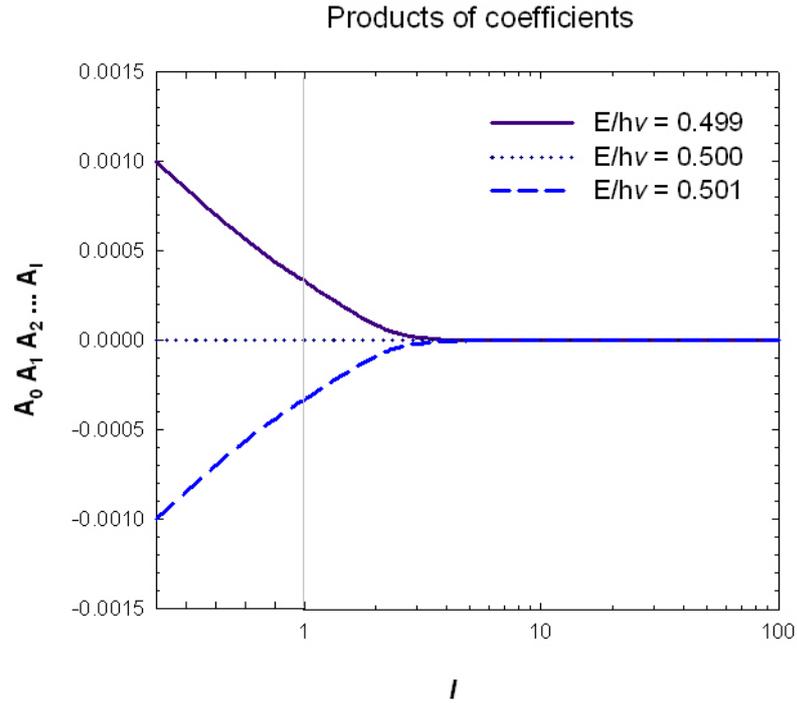

**Fig. 1.** The products of the coefficients $A_\ell$ as a function of $\ell$ for different $E/h\nu$ values, 0.499, 0.500, and 0.501.

On the other hand, it would be nice to show the terms of the power series for different $\ell$. We calculate different terms of the series. Thus, we conclude that, for large $\ell$ and $E/h\nu$=0.499, the last terms go to zero (Fig. 2). So we can stop adding terms in the infinite series where the last calculated terms ($\ell$ >100) are sufficiently small.

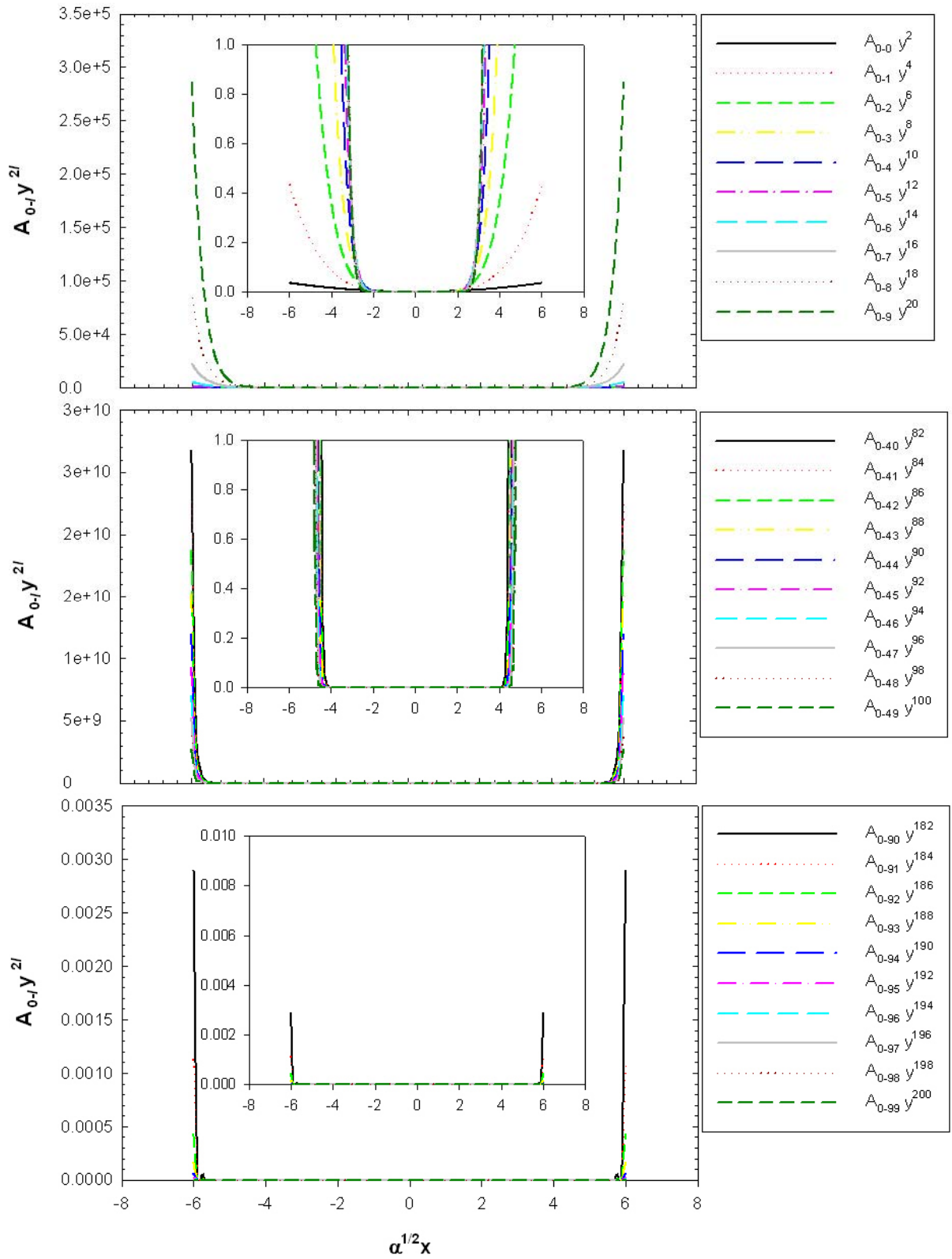

**Fig. 2.** The terms of power series $A_{0-\ell} x^{2\ell}$ as a function of $\alpha^{1/2} x$ for $E/h\nu = 0.499$.

## 4. Results and Discussion

The following figure plots $\psi/c_0$ versus $\alpha^{1/2}x$ for the values of E/h$\nu$=0.499, 0.500, and 0.501, where the recursion relation is used to calculate the ratio $c_{2\ell+2}/c_\ell$. It can be clearly seen in Fig. 3 that the cases of 0.499 and 0.501 go to infinity at a particular value of $\alpha^{1/2}x$. In the region around x=0, the three curves nearly coincide. For $|\alpha^{1/2}x|>3$, the E/h$\nu$=0.500 curve nearly coincides with the x axis, but others does not. In Fig. 4 we also show $[\psi/c_0]^2$ for various stationary states of the harmonic-oscillator. For the solutions of $E/h\nu \pm 0.001$ to any stationary state of Schrödinger equation the numerical values of wave function $\psi/c_0$ or $[\psi/c_0]^2$ are infinite; in that case normalization is going to make it infinite. Such non-normalizable solutions cannot represent particles, and must be rejected.

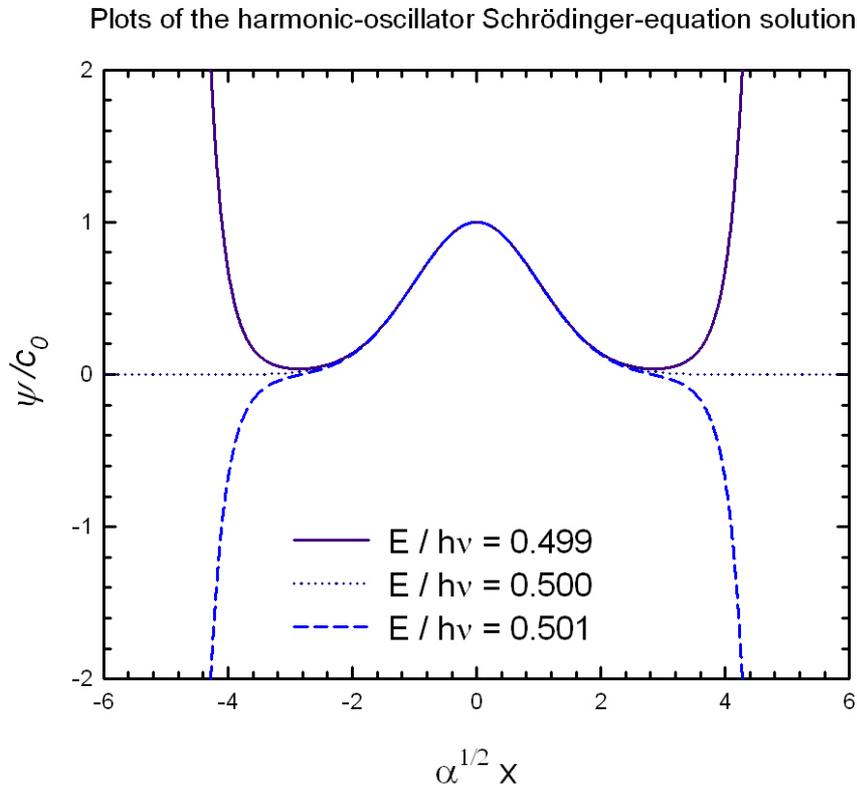

**Fig. 3.** The plots of the harmonic oscillator Schrödinger-equation solution ($\psi/c_0$) as a function of $\alpha^{1/2}x$ for $E/h\nu$ = 0.499, 0.500, and 0.501.

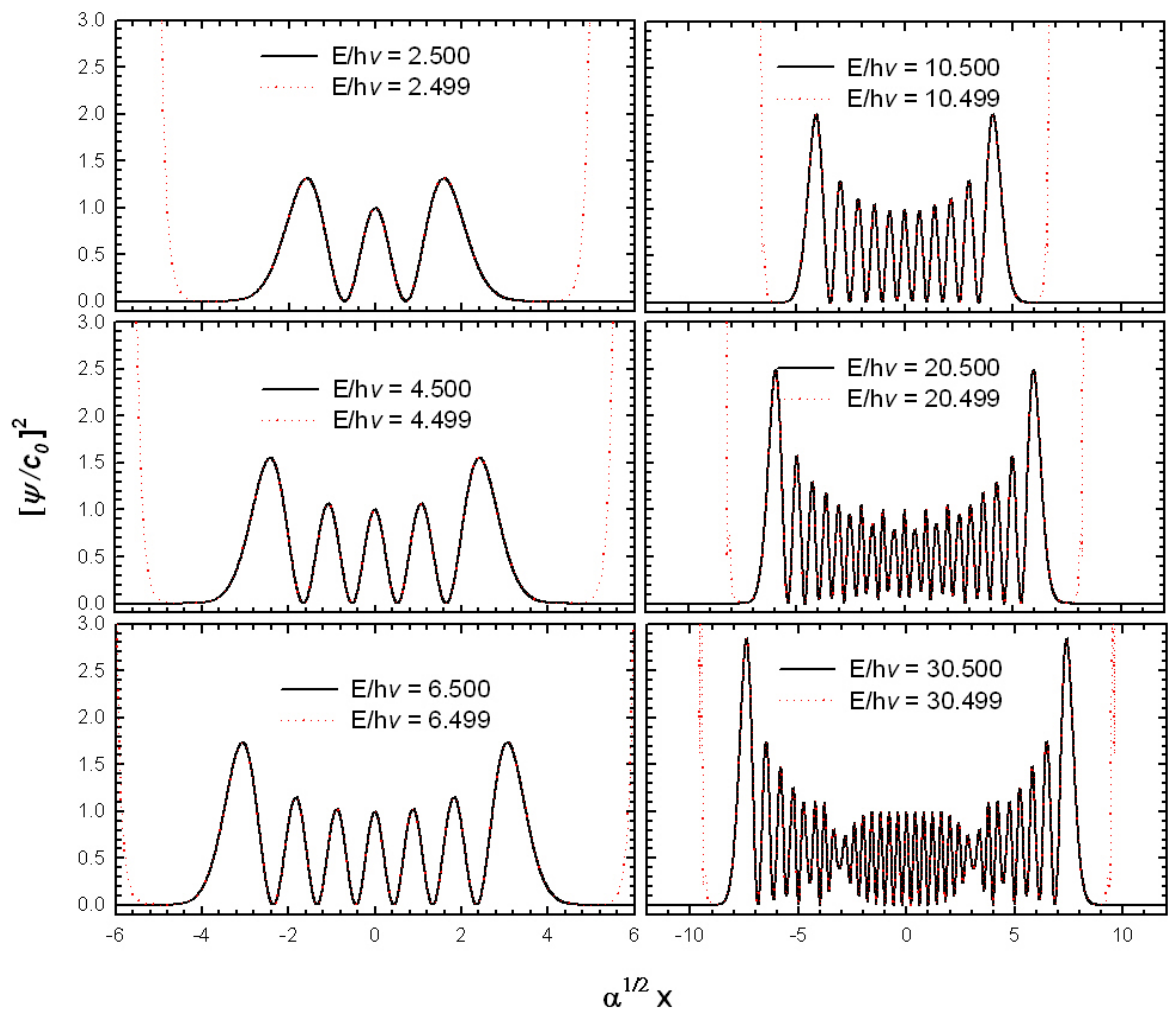

**Fig. 4.** Various stationary states of the harmonic-oscillator. Graphs of $[\psi/c_0]^2$ with $E/h\nu \pm 0.001$ (where $E/h\nu$ =2.5, 4.5, 6.5, 10.5, 20.5, and 30.5)

5. **Conclusion**

We have written a simple program that solves the harmonic-oscillator Schrödinger equation at the level of high school or college physics, and studied the quantization of the energy in the harmonic oscillator potential. It can be shown that the power series solution is a useful way to show the quantization of the energy.